\documentclass[mathleft, final]{an}
\usepackage{graphicx}
\usepackage{times}
\usepackage{natbib}
\bibpunct{(}{)}{;}{a}{}{,}

\renewcommand{\vec}[1]{\mbox{\boldmath $#1$}}

\def \qu{\qquad}
\def \ea{{et\thinspace al.}}

\def \Rey {\mathrm{Re}}

\def \Ha {\mathrm{Ha}}
\def \Pm {\mathrm{Pm}}
\def \q  {\mathrm{q}}
\def \Om  {{\it \Omega}}

\def \A{Alfv\'en}
\def \Ac{Alfv\'enic}
\def \Mm {\mathrm{Mm}}

\begin{document}

\Pagespan{1}{}%
\Yearpublication{2014}%
\Yearsubmission{2014}%
\Month{}%
\Volume{}%
\Issue{}%
\DOI{}%

\title{Axisymmetry vs.\ nonaxisymmetry of a Taylor-Couette flow with \\
\mbox{azimuthal} magnetic fields}
\titlerunning{MHD Taylor-Couette flow with azimuthal fields}

\author{M. Gellert\thanks{Corresponding author: {mgellert@aip.de}} \and  G. R\"udiger}
\authorrunning{M. Gellert \and G. R\"udiger}
\institute{ Leibniz-Institut f\"ur Astrophysik Potsdam (AIP), An der Sternwarte 16, D-14482 Potsdam, Germany}

\received{later}
\accepted{later}
\publonline{later}


\abstract{%
The instability of a supercritical Taylor-Couette flow of a conducting fluid with resting outer cylinder  under the
influence of a uniform axial electric current is investigated for  magnetic Prandtl number $\Pm=1$. In the linear theory the
critical Reynolds number for axisymmetric perturbations  is not influenced by the current-induced axisymmetric magnetic 
field but all  axisymmetric  magnetic perturbations decay. The nonaxisymmetric perturbations with ${m=1}$ are excited even 
without rotation for large enough Hartmann numbers (``Tayler instability''). For slow rotation their growth rates scale with 
the \A\ frequency of the magnetic field but for fast rotation they scale with the rotation rate of the inner cylinder. In the
nonlinear regime  the ratio of the energy of the magnetic $m=1$ modes and the toroidal background field is very low for
the non-rotating Tayler instability but it  strongly grows if differential rotation is present. For super-\Ac\ rotation 
the energies in the ${m=1}$ modes of flow and field  do not depend on the molecular viscosity, they  are almost in 
equipartition and  contain only  1.5\,\% of the centrifugal energy of the inner cylinder. The geometry of the excited 
magnetic field pattern is strictly nonaxisymmetric for slow rotation but it is of the mixed-mode type for fast rotation -- contrary 
to the situation which has been observed at the surface of Ap stars.
}

\maketitle

\section{Introduction}
In recent years instabilities in rotating conducting fluids under the influence of magnetic fields
became of high interest. Especially in view of astrophysical applications  the consideration of
differential rotation is relevant. It is known for a long time that differential rotation with
negative shear (``subrotation'') becomes unstable under the influence of an axial field \citep[]{veli59}. 
Both ingredients of the MagnetoRotational Instability (MRI) itself, i.e. the axial field and also
the not too steep differential rotation, are stable. Nevertheless the full system proves to be unstable. 
One can show for conducting fluids between rotating cylinders that the fundamental mode of the instability is  axisymmetric \citep[]{gellert_2012}
while nonaxisymmetric modes only exist for higher eigenvalues. The opposite is true if the axial
field is replaced by an axial electric current. The toroidal magnetic field due to this (assumed to be 
homogeneous) current is simply ${B_\phi\propto R}$ (with $R$ as the distance from the rotation
axis). Such a profile is unstable against nonaxisymmetric modes as the necessary and sufficient
criterion for stability against nonaxisymmetric perturbations,
\begin{equation}\label{eq1}
 \frac{{\rm d}}{{\rm d}R}(R B^2_\phi)<0 
\end{equation} 
\citep[]{tayler_1973}, is {\em not} fulfilled. The field profile caused by a homogeneous current can 
thus be expected to develop {\em nonaxisymmetric} magnetic field patterns. The existence of just this 
instability has been demonstrated by a recent experiment in the laboratory with liquid metal (Seilmayer et al. 2012).

The criterion of stability against {\em axisymmetric} perturbations under the presence of differential rotation and 
for the same toroidal field,
\begin{equation}\label{micha}
 \frac{1}{R^3}\frac{{\rm{d}}}{{\rm{d}}R}(R^2\Om)^2 > 0 
\end{equation}
\cite[]{michael_1954}, is identical with  the corresponding hydrodynamic formulation. Thus the question is 
whether the relation (\ref{micha}) also ensures the stability not only of a hydrodynamic axisymmetric disturbance 
but also the stability of an axisymmetric magnetic disturbance. We shall find that for the Taylor-Couette (TC) flow 
with resting outer cylinder the magnetic axisymmetric mode only behaves passively under the influence of the 
axisymmetric background field. The axisymmetric perturbation mode, therefore, grows for supercritical Reynolds 
number only by the influence of the neighbor modes. The kinetic axisymmetric mode, however, behaves linearly 
unstable for supercritical rotation of the inner cylinder.

There are many open questions about the kinetic and magnetic energies of the nonaxisymmetric modes. The main question 
is whether large shear of the fluid supports or reduces their energy. On the one hand it is obvious that strong 
differential rotation should lead to a suppression of the magnetic energy as nonuniform rotation for high enough 
electric conductivity always suppresses nonaxisymmetric modes. On the other hand, weak differential rotation supports 
the excitation of kink-type modes in contrast to rigid rotation. A TC flow with resting outer cylinder may easily 
serve as a model to answer such questions. The Reynolds number of the rotation of the inner cylinder is the only 
remaining free parameter. For this model the energies can be normalized  with the centrifugal energy $R^2 \Om^2$ 
taken from the inner cylinder. The same formulation can also be used for the hydrodynamic TC flow without magnetic 
field and the results can be compared. The question is  how much centrifugal energy is stored in the nonaxisymmetric 
flow and field modes. And how is (for given magnetic field) the dependence of these energies on the Reynolds number 
and the magnetic Prandtl number?  We shall show that for fast rotation the resulting 
values $\lim \hat  \q$ for $\Om\to \infty$ in the relations
\begin{equation}\label{lim}
 \langle {\vec{u}}^2\rangle ={\hat \q}_{\rm kin} R_{\rm in}^2\Om_{\rm in}^2, \ \ \ \ \ \  \langle {\vec{b}}^2\rangle =\mu_0\rho\,{\hat \q}_{\rm mag} R_{\rm in}^2\Om_{\rm in}^2
\end{equation}
do {\em not} depend on the Reynolds number. Hence, the molecular viscosity of the fluid does {\em not} determine 
the turbulent energies. This is also true for the hydrodynamic TC flow. With other words, the kinetic and the 
magnetic energies saturate for the limit $\nu\to 0$. One might assume that for $\Om\to\infty$ the influence of 
the magnetic field vanishes so that $\lim \hat \q_{\rm kin}$ becomes equal to the value of the hydrodynamical 
flow. In this case the question remains how the associated magnetic energy behaves. It would be suggestive to 
assume that $\lim \hat \q_{\rm mag}$ also vanishes for $B\neq 0$ and for very fast rotation -- but this is not the case.

\section{The model}\label{sec_model}
The most simple model to study the interaction of differential rotation and Tayler instability  is
the classical TC system. A Reynolds number may represent the rotation of
the inner cylinder and forms the only free parameter of rotation if the outer cylinder is at rest.  

The equations to describe the problem are the MHD equations
\begin{eqnarray}\label{mhd2}
 \frac{\partial \vec{U}}{\partial t} + (\vec{U}\cdot \nabla)\vec{U} = -\frac{1}{\rho} \nabla P + \nu \Delta \vec{U} 
   + \frac{1}{\mu_0\rho}{\textrm{curl}}\vec{B}\! \times\! \vec{B}
\end{eqnarray}
and 
\begin{eqnarray}\label{mhd22}
 \frac{\partial \vec{B}}{\partial t}= {\textrm{curl}} (\vec{U} \!\times\! \vec{B}) + \eta \Delta\vec{B},  
\end{eqnarray}
 with  ${\textrm{div}}\ \vec{U} = {\textrm{div}}\ \vec{B} = 0$  for an incompressible fluid, 
where $\vec{U}$ is the velocity, $\vec{B}$ the magnetic field, $P$ the 
pressure, $\nu$ the kinematic viscosity, and $\eta$ the magnetic diffusivity.

The  basic state in the cylindrical system $(R,\phi,z)$ is $ U_R=U_z=B_R=B_z=0$ and 
\begin{eqnarray}\label{basic}
 B_\phi = \frac{B_{\rm in}}{R_{\rm in}} R, \qu U_\phi = \Om R = a R + \frac{b}{R}, 
\end{eqnarray}
where $a$ and  $b$ are constant values defined by 
\begin{eqnarray}\label{ab}
 a=-\frac{r_{\rm in}^2\Om_{\rm in}}{1-r_{\rm in}^2}, \qu b=
  \frac{\Om_{\rm in} R_{\rm in}^2}{1-r_{\rm in}^2}.
\end{eqnarray}
Here is $r_{\rm in}={R_{\rm in}}/{R_{\rm out}}$, where $R_{\rm in}$ and $R_{\rm out}$ are the radii, $\Om_{\rm in}$  and $B_{\rm in}$ the angular velocity 
and the azimuthal magnetic field of the inner cylinder, respectively. The radial magnetic profile is the profile of an applied homogeneous axial electric current 
\cite[see][]{tayler_1957}. Throughout the whole paper, the radius ratio is always fixed to $r_{\rm in}=0.5$.

The dimensionless physical parameters defining the problem are the magnetic Prandtl number $\Pm$, the 
Hartmann number $\Ha$, and the Reynolds number $\Rey$, i.e. 
\begin{eqnarray}\label{pm}
 {\Pm} = \frac{\nu}{\eta}, \qu
 {\Ha} =\frac{B_{\rm in} D}{\sqrt{\mu_0\rho\nu\eta}},  \qu
 {\Rey} =\frac{\Om_{\rm in} D^2}{\nu},
\end{eqnarray}
where the gap width ${D=R_{\rm out}-R_{\rm in}}$ is used as unit of length. 
The \A\ frequency 
\begin{equation}\label{OmA}
\Om_{\rm A}=\frac{B_{\rm in}}{\sqrt{\mu_0\rho D^2}}
\end{equation}
in relation to the angular velocity $\Om_{\rm in}$ indicates whether the system is slowly or rapidly rotating.
Through the whole paper the magnetic Prandtl number is fixed to unity, $\Pm=1$. In this case the magnetic Mach number
\begin{equation}\label{Mm}
\Mm= \frac{ \Om_{\rm in} }{\Om_{\rm A}}
\end{equation}
which indicates slow ($\Mm<1$) and fast ($\Mm>1$) rotation simply equals $\Rey/\Ha$.

To  solve the equations a spectral element code has been  used which is based   on the hydrodynamic code 
of \cite{fournier_2005}. It works with an expansion of the solution in Fourier modes in the azimuthal 
direction. The remaining part is a collection of meridional problems,
each of which is solved using a Legendre spectral element method \cite[see, e.g.,][]{dev_2002}.
Between $8$ and $16$ Fourier modes are used. The polynomial order is varied between $10$ and
$16$ with two elements in radial direction. The number of elements in axial direction corresponds to
the aspect ratio $\Gamma$, the height of the numerical domain  in units of the gap width, thus the spatial 
resolution is the same as for the radial 
direction. With a semi-implicit approach consisting of second-order backward differentiation  and 
third order Adams-Bashforth for the nonlinear forcing terms time stepping is done with second-order 
accuracy. Periodic conditions in axial direction are applied to minimize finite size effects. With $\Gamma=8$ 
all excitable modes in the analyzed parameter region fit into the system. The boundary conditions at 
the cylinder walls are always assumed to be no-slipping and the cylinders are considered as perfect conductors.
All linear solutions are optimized so that the searched for wave number $k$ yields the lowest Reynolds number. 

\begin{figure}
\includegraphics[width=0.48\textwidth,height=0.22\textheight]{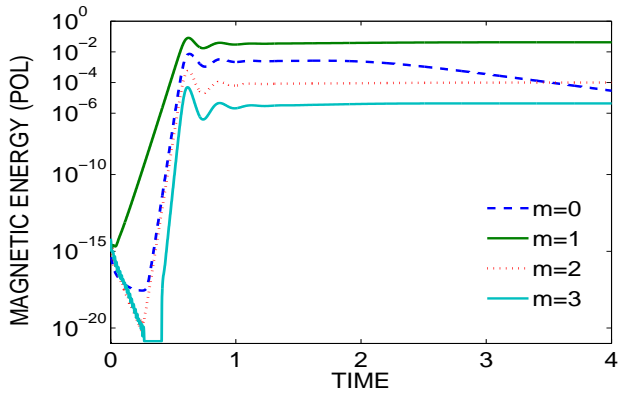} 
\caption{The energy of the poloidal magnetic components  (normalized with the energy of $B_{\rm in}$) of the modes with
         $m=0,1,2$. The time is measured in diffusion times, $D^2/\eta$, both cylinders are resting. Only the mode with 
         $m=1$ is linearly unstable. The field components possess   initial amplitudes of $10^{-8}$.  $\Ha=80$, $\Rey=0$.} 
\label{fig1a} 
\end{figure}

\cite{tayler_1957} found that magnetohydrodynamically all perturbations of the axial current with an azimuthal 
wave number ${m=1}$ are unstable. One can show that the system is degenerated under the transformation $m\to -m$ so that all
eigenvalues (for specific  $\Ha$) are simultaneously valid for each pair $m=\pm 1$ and $ \pm 2, \ldots$ .
The critical Hartmann number $\Ha=35.3$ for the  $m=\pm 1$ mode and  for $r_{\rm in}=0.5$ does {\em not} depend on the
magnetic Prandtl number \cite[]{rued_2010}. Figure \ref{fig1a} concerns  a numerical 
realization of the instability with $\Ha=80$. The plot shows the exponential growth and the saturation of the 
energy $\langle b^2_R+b^2_z \rangle/B^2_{\rm in}$. The azimuthal component is not included as it exceptionally grows 
as a result of the saturation process. One finds that indeed only the mode with $m=1$ is linearly 
unstable while the neighboring modes with $m=0, 2, 3$  are nonlinearly coupled to the
instability. Without rotation the instability of an axial electric current is of the kink-type,
thus the axisymmetric mode does not play an important role (see Fig. \ref{fig3g}, left). Note that 
for $\Rey=0$ the whole pattern does  {\em not} drift in the azimuthal direction.

\section{Growth rates and instability pattern}
\begin{figure}
\includegraphics[width=0.48\textwidth,height=0.22\textheight]{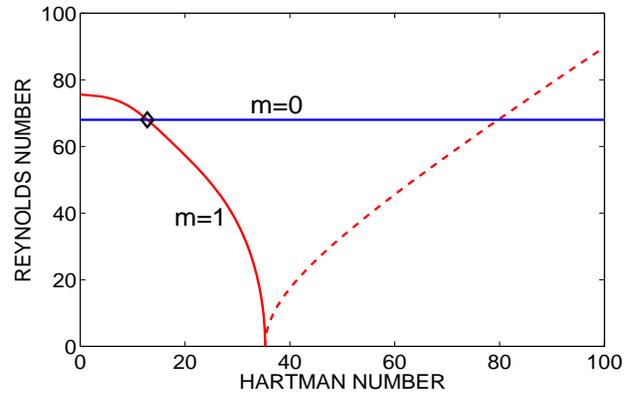} 
\caption{The instability map for a rotating container for rigid rotation (dashed) and for resting outer cylinder (solid). 
         The rhombus marks the transition from the excitation of axisymmetric pattern to nonaxisymmetric pattern. The 
         critical Hartmann number at the horizontal axis ($\Ha=35.3$) holds for all $\Pm$. }
\label{fig2a}
\end{figure}

The inner cylinder may now rotate with the rotation frequency $\Om_{\rm in}$ while the  outer cylinder rests so that the rotation 
law is rather steep. For this case Fig. \ref{fig2a} gives the lines of marginal excitation derived from the linearized equations. 
The solid lines describe the instability curves for the modes ${m=0}$ and ${m=1}$ while for comparison the dashed line marks the curve 
of marginal Tayler instability for rigid container rotation. Note that the solid line marked with $m=0$ does not show any magnetic 
influence. Axisymmetric Taylor vortices  for this configuration are unstable for $\Rey\geq 68$ with and  without magnetic field. It 
is not yet clear, however, whether the condition $\Rey> 68$ for supercritical excitation of axisymmetric modes concerns the flow 
pattern, the field pattern or even both.

In the instability map for the wavy mode with $m=1$ the hydrodynamical instability (without
magnetic field) with $\Rey=76$  and the critical magnetic field for the Tayler instability without rotation with 
$\Ha=35$ are directly  connected. If only the lowest Reynolds numbers are considered then a transition
exists from axisymmetric perturbations for low $\Ha$ to nonaxisymmetric perturbations for high
$\Ha$ which is marked by a rhombus in Fig. \ref{fig2a}.

\begin{figure} 
\includegraphics[width=0.48\textwidth]{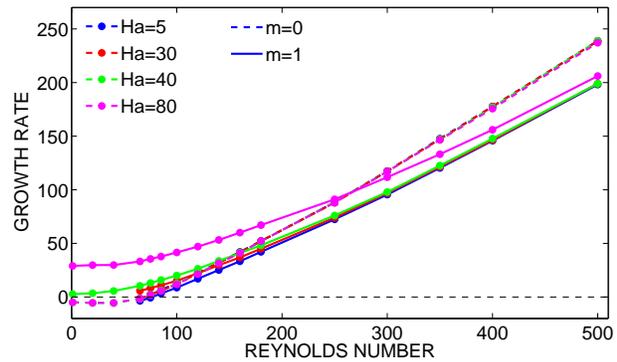} 
\caption{Growth rates of the {\em flow} normalized with the diffusion frequency for the 
         modes ${m=0}$ and ${m=1}$. The solid curves for the ${m=1}$ mode are also valid for 
         the magnetic field disturbances. The  axisymmetric kinetic mode (dashed line, for all $\Ha$) is unstable in 
         contrast to the axisymmetric magnetic mode.} 
\label{fig2} 
\end{figure}

Figure \ref{fig2} gives the  growth rates  $\omega_{\rm gr}$ for the exponential growth of unstable disturbances  of the
kinetic modes. They are normalized with the viscosity frequency. The dashed lines denote the axisymmetric flow mode for 
various Hartmann numbers. They are identical for all applied field strengths so that the magnetic influence vanishes. 
This pure hydrodynamical axisymmetric mode leads to $\Rey=68$, and is {\em not} suppressed by the toroidal magnetic
field. 

The growth rates of the nonaxisymmetric modes with $m=1$ in Fig. \ref{fig2} (solid lines) behave completely different. 
For sufficiently strong magnetic fields  they become unstable even without global rotation. Without rotation the growth 
rate runs with  the \A\ velocity $\Om_{\rm A}$. For slow rotation the growth rates depend on both the Hartmann number 
and the Reynolds number while for fast rotation the Hartmann number dependence disappears. In the latter case it is 
$\omega_{\rm gr}\propto \Om_{\rm in}$ which for fast rotation exceeds  $\Om_{\rm A}$. One finds that the differential 
rotation leads to a rotational amplification of the instability rather than a rotational suppression as it happens for 
rigid rotation. For ${\Mm \gg 1}$ the  current-driven nonaxisymmetric modes under the presence 
of differential rotation result as rotationally supported.

The growth rates of the  nonaxisymmetric modes in Fig.\,\ref{fig2} are identical for kinetic and magnetic modes.
One finds, however, that the growth rate for the axisymmetric magnetic mode vanishes for all $\Rey$ and $\Ha$. 
The axisymmetric magnetic mode remains uninfluenced and does not become unstable. This is not unexpected as in 
the linear theory its equation decouples from the entire system of equations (R\"udiger et al. 2011).

\begin{figure}
\includegraphics[width=0.48\textwidth]{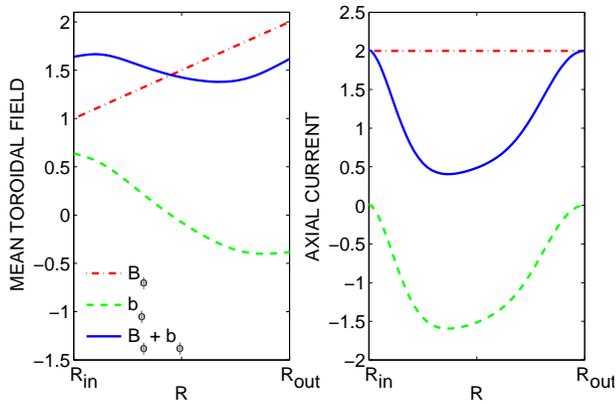}
\caption{\emph{Left panel}: The induced axisymmetric magnetic mode (dashed) reduces the azimuthal magnetic background field (dot-dashed) producing the 
         almost uniform azimuthal field (solid). \emph{Right panel}: the same for the corresponding electric currents. The quantities are averaged over 
         $\phi$ and $z$. $\Ha=80$, $\Rey=150$.}
\label{fig4b}
\end{figure}

In the nonlinear simulations the toroidal magnetic $m=0$ mode behaves different. The stronger the toroidal 
background field the more energy is stored in the $m=0$ mode of the toroidal instability pattern. For the 
particular example with ${\Ha=80}$ and ${\Rey=150}$, Fig. \ref{fig4b} demonstrates  that the sign of the axial 
electric current which is formed by the axisymmetric magnetic instability mode is opposite to the sign of 
the background current. The dot-dashed lines in Fig. \ref{fig4b} symbolizes the toroidal background field 
(left panel) and the uniform axial background current (right panel). This current is reduced by the negative 
axial current induced by the instability (dashed  line). In the average the resulting total electric current 
(solid line) is about 50\% of the undisturbed background current. The eddy magnetic-diffusivity in the 
saturated state is thus just of the order of the molecular magnetic-diffusivity (Gellert \& R\"udiger 2009). 
Obviously, the system saturates by  the effective reduction of the applied electric current, or, what is the 
same, by the increase of the effective magnetic diffusivity. 

\begin{figure}
\includegraphics[width=0.24\textwidth]{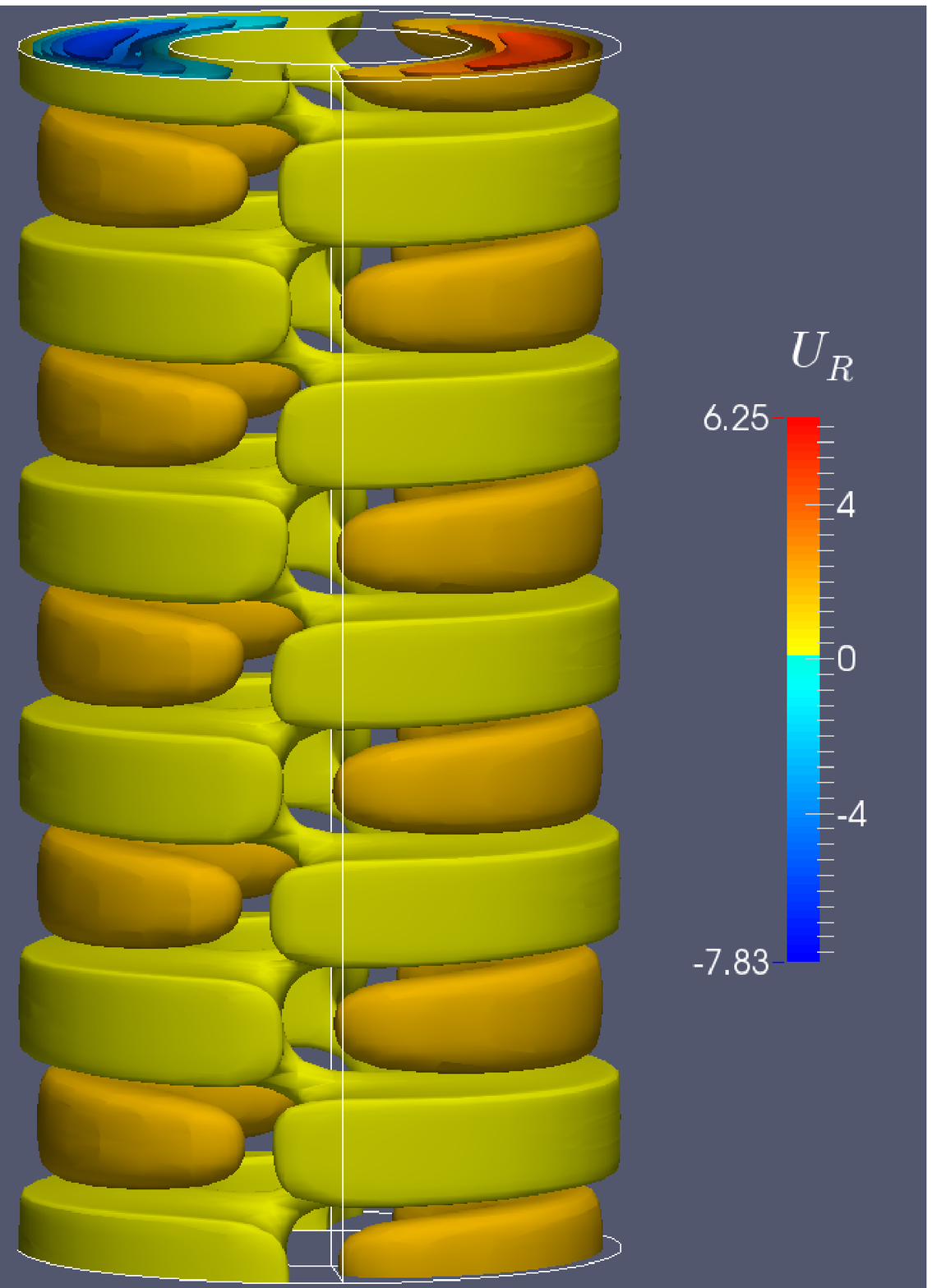}
\hfill
\includegraphics[width=0.24\textwidth]{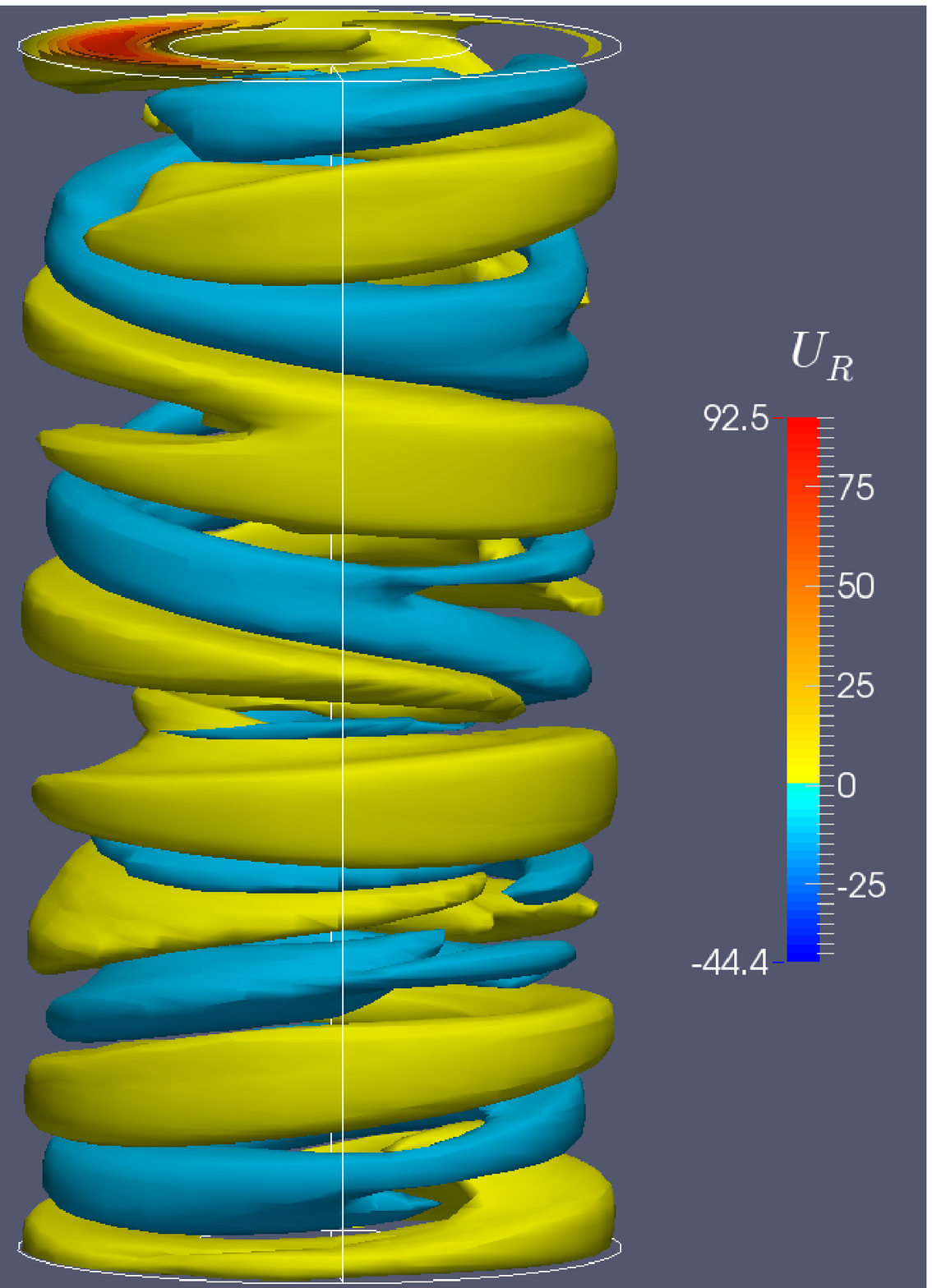}
\caption{The radial flow $u_R$  measured in units of the diffusion velocity for resting container (\emph{left panel}) and 
         for $\Rey=350$ (\emph{right panel}). Only the instability pattern of rotating fluids is drifting in the positive 
         azimuthal direction. $\Ha=80$.}
\label{fig3g}
\end{figure}

In Fig. \ref{fig3g} the patterns of the radial flow induced by the instability are compared for the cases of resting inner cylinder 
(left) and rotating inner cylinder (right, $\Rey=350$). For the resting container  also the pattern is resting while it drifts in the 
positive azimuthal direction in the rotating case. The flow velocities are given in units of the viscosity velocity $ \eta/D$ which 
must be divided by $\Rey$ to get the stream flow in units of the linear velocity of the rotating cylinder. Note that the $m=1$ mode 
less dominates the patterns if the inner cylinder rotates. The rotation favours the development of instability patterns which are of 
the mixed-mode type. The axisymmetric mode, however,  never dominates the nonaxisymmetric modes.

The vertical wave number differs only slightly for the two examples.

\section{The energies}\label{sect_nonl}
Following the linear results, the instability pattern should be axisymmetric in the flow system but nonaxisymmetric in the 
field system. If this would be the final truth then a magnetic configuration with mainly axisymmetric geometry hardly results 
under the presence of differential rotation. However, the simulations demonstrate also the nonlinear interaction of the modes. 
After reaching a saturated state the energies in the axisymmetric and nonaxisymmetric modes are almost 
equal. 
\begin{figure}
\includegraphics[width=0.48\textwidth]{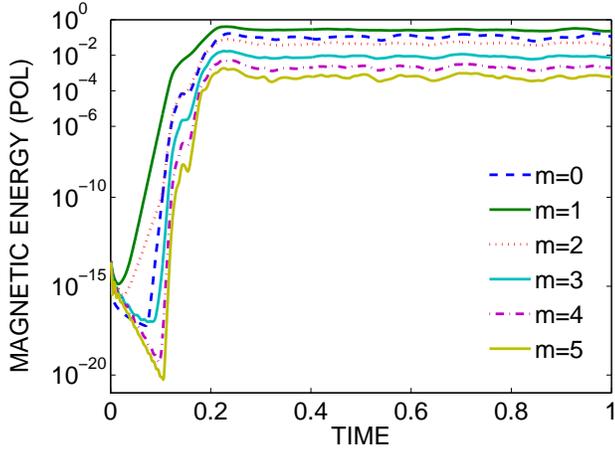}
\caption{The same as in Fig. \ref{fig1a} but for $\Rey=350$. }
\label{fig3b}
\end{figure}

The Figs. \ref{fig1a} and \ref{fig3b}  present the modal  structure of the saturated instability pattern in detail for the 
energy of the poloidal magnetic components. As shown in Fig. \ref{fig3b} the energy of the mode $m=1$ hardly  dominates 
the energy of the mode $m=0$. The modes which are stable in the linear theory decay at the beginning of the simulation but 
later they receive energy from the unstable $m=1$ mode. The growth time for the magnetic $m=1$ mode taken from the 
Fig. \ref{fig2} is only 0.4 rotation times in correspondence to the temporal behavior in Fig. \ref{fig3b}. 
The plots also suggest that with (differential) rotation the excitation of the unstable and stable modes is much 
faster than without (differential) rotation and the saturation levels are higher. The dashed line in Fig. \ref{fig2a}
demonstrates that the situation for rigid rotation is rather different.

The enslaved neighboring modes ${m=0}$ and ${m=2}$ possess nearly the same energy. It is demonstrated that the difference 
to the $m=1$ mode is reduced by faster rotation. It is thus the rotation which transfers energy into the stable 
modes. The relative energy of the axisymmetric mode for $\Mm=0$ is much smaller (see Fig. \ref{fig1a}).
This is insofar unexpected as differential rotation usually leads to a smoothing of nonaxisymmetric (poloidal) magnetic fields. 
Note that the axis of the magnetic field pattern of Ap stars is {\em not} orthogonal to the axis of rotation (Oetken 1977). 
The ratio of nonaxisymmetric to axisymmetric field parts is larger for fast rotation (Landstreet \& Mathys 2000). Our results 
do {\em not} comply with these observations of magnetic stars. One must be careful, however, as the Reynolds numbers in the present 
paper do only concern to the inner rotation rate rather than to the observed rotation of the stellar surface.

In the following the kinetic and magnetic energies which are stored in the nonaxisymmetric modes of the instability  are presented. 
Because of its dominance it is allowed to concentrate to the modes with ${m=1}$. We denote  the nonaxisymmetric flow perturbations 
with $\vec{u}$ and the field perturbations with $\vec{b}$. One finds that the ratio 
\begin{equation}\label{eps}
\epsilon= \frac{ \langle\vec{b}^2\rangle }{\mu_0\rho\langle\vec{u}^2\rangle}
\end{equation}
of the magnetic and the kinetic energy for fast rotation  with ${\Mm>1} $ is of order unity. For fast rotation the
instability is thus {\em not} dominated by the magnetic fields.  The simulations show that for the nonaxisymmetric modes the 
magnetic energy only dominates the kinetic energy for slow rotation and/or strong fields (Fig. \ref{fig3b}). 
\begin{figure}
\vskip-2mm
\includegraphics[width=0.48\textwidth,height=0.22\textheight]{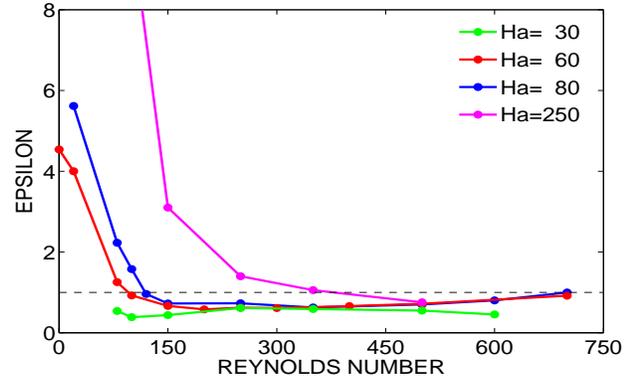}
\caption{The ratio $\epsilon$ of the energy of the nonaxisymmetric  magnetic modes to the energy of the  kinetic modes for $m=1$.}
\label{fig3ab}
\end{figure}

\begin{figure}
\includegraphics[width=0.48\textwidth,height=0.22\textheight]{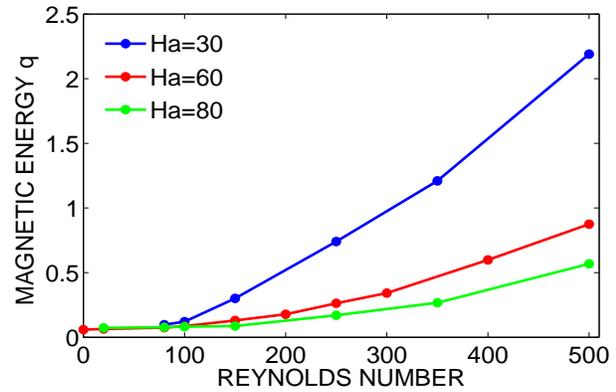}
\caption{The quantity $\q$ for the magnetic modes with $m=1$. }
\label{fig3d}
\end{figure}

We find, on the other hand, that for slow rotation (${\Mm<1} $) the magnetic energy of the instability 
perturbations always exceeds their kinetic energy. In Fig. \ref{fig3d} the dimensionless magnetic energy ratio
\begin{equation}\label{q1}
\q=\frac{\langle \vec{b}^2 \rangle}{B_{\rm in}^2}
\end{equation}
between the energy of the fluctuations of the magnetic mode with the azimuthal number $m=1$ and the toroidal magnetic
background field is plotted. The square-root of $\q$ yields the ratio of the rms value of the magnetic fluctuation 
to the magnetic background field.

\begin{figure}
\vskip-1mm
\includegraphics[width=0.48\textwidth,height=0.22\textheight]{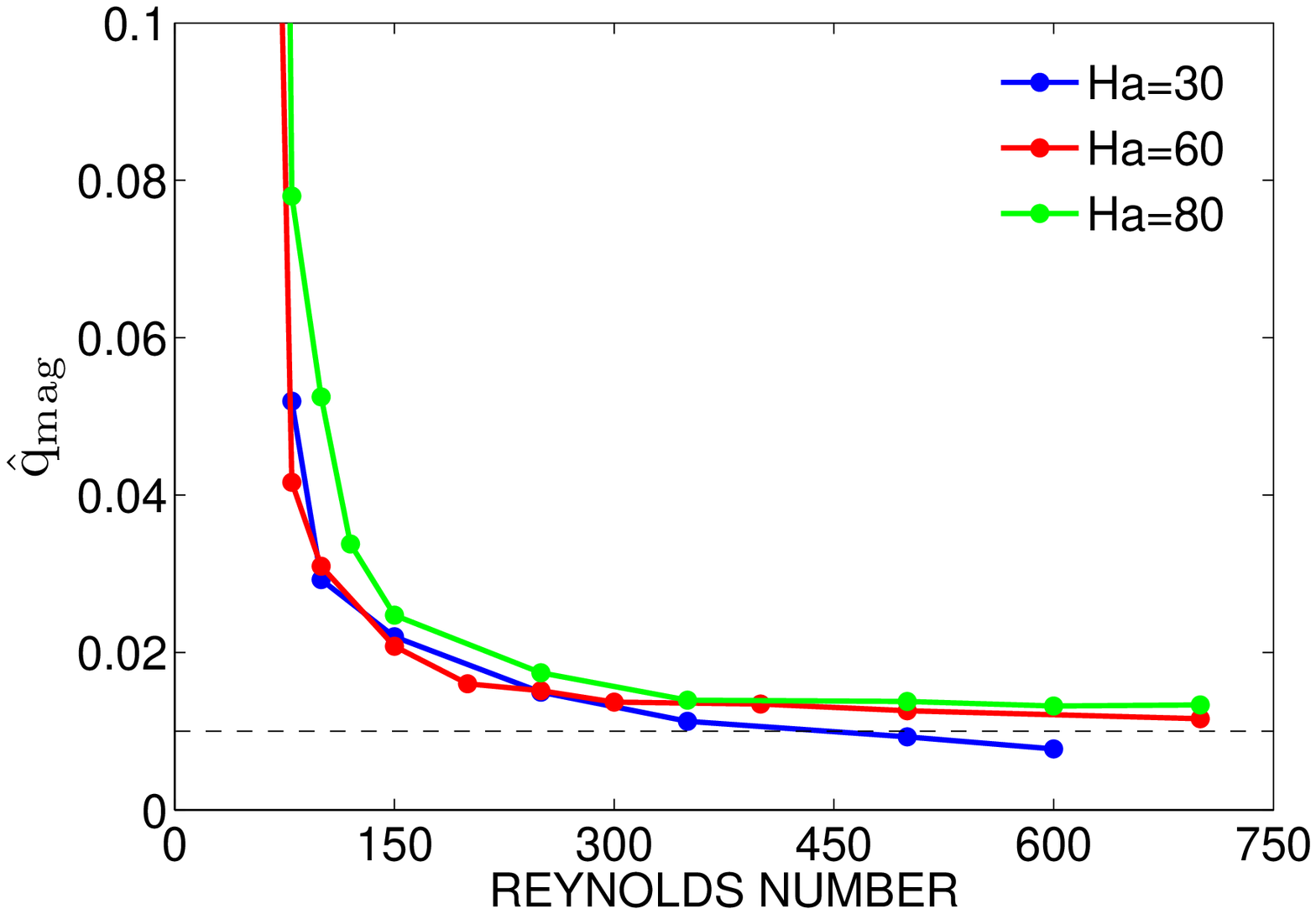}
\vspace{1mm}
\includegraphics[width=0.48\textwidth,height=0.22\textheight]{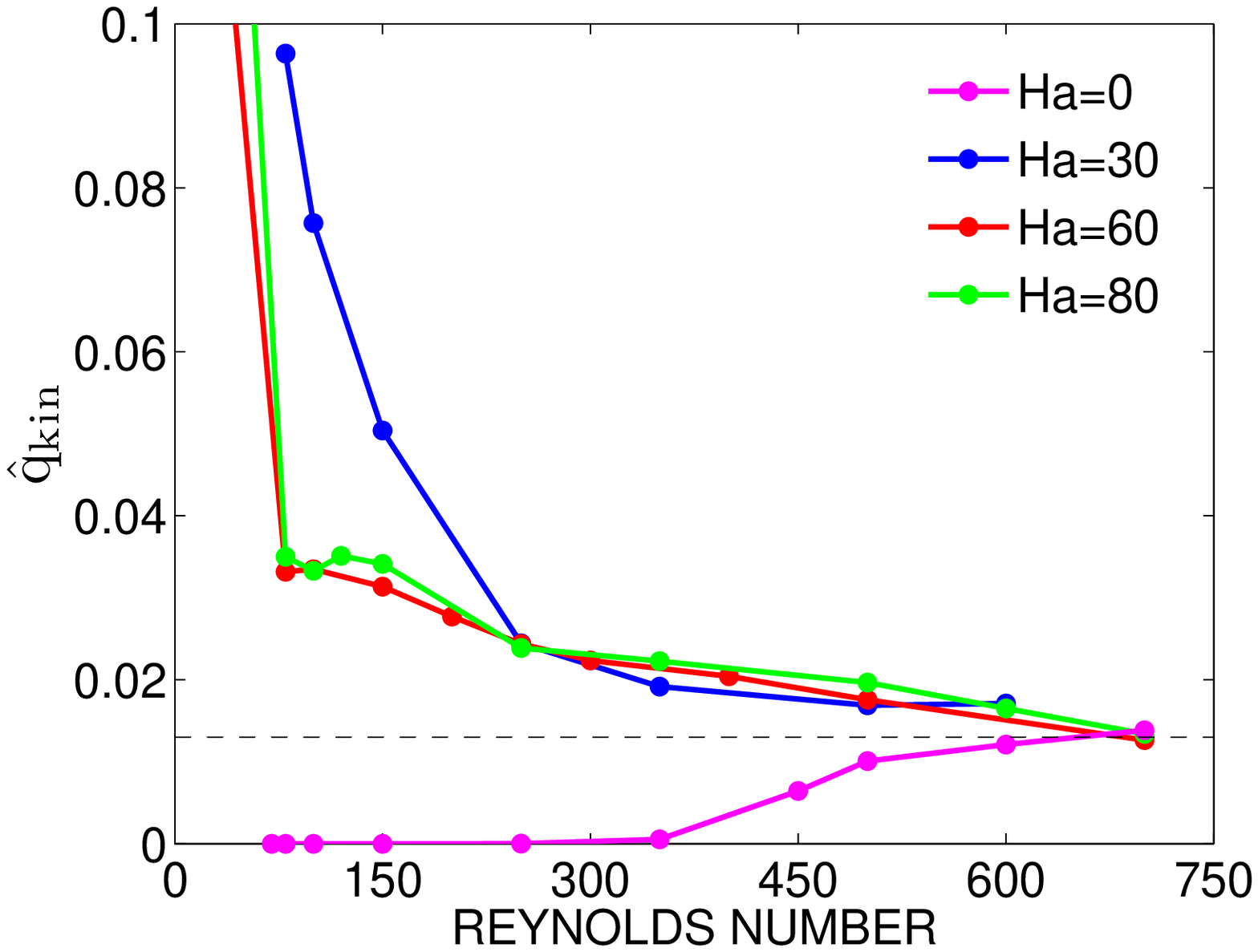}
\caption{The same as in Fig. \ref{fig3d} but for the quantity $\hat{\q}$ of the magnetic 
(\emph{top}) and the kinetic (\emph{bottom}) 
         energy. Note that for fast rotation the influence of the Reynolds number disappears so that  the molecular 
         diffusivities do not influence the numerical values of both the energies. This is also true for the kinetic 
         energy of the $m=1$ mode of the hydrodynamic TC flow (bottom plot, $\Ha=0$). }
\label{fig3e}
\end{figure}

Without rotation the instability provides a small basal value $\q$ of about 0.05. Increasing rotation lets the 
$\q$ grow which, however, sinks for growing $\Ha$. For fast rotation the profiles in Fig. \ref{fig3d} suggest the
approximative behavior of the quantity ${\hat{\q}}={\Ha}^2 {\q}/ {\Rey}^2$ or what  for $\Pm=1$ is the same, 
\begin{equation}\label{q111}
{\hat{\q}= \frac{\Om_{\rm A}^2}{\Om_{\rm in}^2}\ {\q} }.
\end{equation}
From test calculations we know that the latter formulation  also holds for $\Pm \neq 1$. The final expression  is thus
\begin{equation}\label{result}
 \frac{\langle \vec{b}^2 \rangle}{\mu_0\rho} = {\hat{\q}_{\rm mag}} \  \Om_{\rm in}^2 D^2
\end{equation}
with the numerical value ${\hat{\q}_{\rm mag}\simeq 0.012}$ for very fast rotation (Fig. \ref{fig3e}, top).
Figure  \ref{fig3e} (bottom) shows that a similar expression, i.e.
\begin{equation}\label{result2}
{\langle \vec{u}^2 \rangle} = {\hat{\q}_{\rm kin}} \  \Om_{\rm in}^2 D^2,
\end{equation}
also holds for the kinetic energy. For ${\Mm>1}$ the coefficients ${\hat{\q}_{\rm kin}}$ and ${\hat{\q}_{\rm mag}}$ 
do not depend on the Reynolds number. Both energies can thus be expressed by the global quantities $R_{\rm in}$ and $\Om_{\rm in}$
and they  are almost in equipartition. The faster the linear rotation of the inner cylinder the 
more energy is stored in the nonaxisymmetric mode of the instability.  The background field obviously only acts as a catalyst which does not
influence the numerical value of the resulting magnetic energy of the instability.

Figure \ref{fig3e} (lower panel) also shows the kinetic energy of the ${m=1}$ mode for hydrodynamic TC flow. For slow rotation this energy is very 
small. For more rapid rotation it linearly grows with the  Reynolds number. Also for nonmagnetic fluids the value of 
$\hat \q_{\rm kin}$ for very fast rotation does not depend on the Reynolds numbers. Moreover, the hydrodynamic model reaches the 
same value of $\hat \q$ as the magnetohydrodynamic model for fast rotation. The basic difference to the magnetohydrodynamic models 
is that for them the kinetic energy for slow and medium rotation becomes smaller for growing Reynolds number contrary to the 
hydrodynamic case. For very large Reynolds number, the kinetic energy in the $m=1$ mode is {\em the same for hydrodynamic and 
magnetohydrodynamic flows}.

The results for super-\A\ rotation ($\Mm>1$) are correct for the energy in ${m=0}$ and ${m=1}$ of the poloidal
magnetic field  and for the $m=1$ mode of the toroidal field. In all these cases the energy contained in the modes -- after saturation -- 
does not exceed about 1.5\,\% of the centrifugal energy of the inner cylinder. One also recognizes that the energy of the
toroidal $m=1$ mode exceeds the energy of both  modes ($m=0$ and $m=1$) of the poloidal components maximally by a factor of two.

\section{Summary}
The stability of the simplest hydromagnetic Taylor-Couette flow system with a resting outer cylinder  has been considered as 
the basic  model  of the interaction of differential rotation and stellar toroidal background fields. The background field is thought 
to be due to a uniform axial electric current, which by itself becomes unstable if its Hartmann number exceeds the value 
${\Ha=35}$ (for a TC container with ${R_{\rm out}= 2 R_{\rm in}}$). In the hydrodynamic regime this rotation law becomes unstable 
for a Reynolds number of the inner cylinder exceeding  ${\Rey=68}$.  While at this threshold value an axisymmetric instability 
pattern is excited, the  current-driven instability without rotation excites a strictly nonaxisymmetric pattern. 
The relation between axisymmetric and nonaxisymmetric instability modes and their kinetic and magnetic energy 
is the main focus of the present paper. 

The growth rates of the linear theory show basic differences for  the modes. One finds the growth rates of the kinetic modes 
increasing with increasing rotation frequency of the inner cylinder. The nonuniform rotation in the container does {\em not} 
suppress the instability -- as rigid rotation does -- but it strongly destabilizes the system. This is also true for the 
nonaxisymmetric magnetic modes with $m= 1$, but the axisymmetric  magnetic mode remains stable. 

That the axisymmetric magnetic mode proves to be linearly stable does not mean, however,  that the resulting pattern in the nonlinear 
regime is fully nonaxisymmetric. We have shown with nonlinear numerical simulations that energy of the unstable 
$m=1$ mode is distributed into the neighboring modes $m=0$ and $m=2,3, \dots $ . In all cases, however, the azimuthal magnetic 
spectrum peaks at $m=1$. For fast rotation this peak energy is exclusively determined by the rotation speed of the inner 
cylinder rather than its Reynolds number. Hence, the amplitude of the magnetic background field, the microscopic viscosity and the 
electric conductivity  do not influence the peak energies. The same is true for the kinetic energy of the nonaxisymmetric modes 
(see Fig. \ref{fig3e}, lower panel). 

The ratio $\epsilon$ of the magnetic and the kinetic energy determines the hydromagnetic  character of the instability. 
Figure~\ref{fig3ab} demonstrates that ${\epsilon>1}$ for supercritical fields with $\Ha>\Ha_{\rm crit}$ only appears for small Reynolds number. For
fast rotation the instability is not magnetic-dominated, the kinetic energy in the MHD turbulence 
cannot be considered as small compared to the magnetic energy. This result should have severe astrophysical
consequences. For fast rotation or weak field the magnetic-induced angular momentum transport does not exist without a 
turbulent mixing of chemicals with the similar intensity, which should strongly affect the stellar structure and evolution. 
On the other hand, for slow rotation  the magnetic-induced eddy viscosity strongly exceeds the mixing 
coefficient (which is not influenced by the magnetic fluctuations) so that indeed in this case the angular momentum can migrate
outwards without any direct influence on the stellar evolution.

Also, in the limit of fast rotation the kinetic energy of the magnetically supercritical system and the kinetic energy of 
the purely hydrodynamical TC flow are almost equal and do not depend on the microscopic viscosity.

The simulations also reveal details about the saturation process of the instability. The system generates an axisymmetric part 
of the toroidal magnetic field which is equivalent to an electric current in opposite direction of the applied current. The 
amplitude of this counter-current corresponds to an increase of the effective magnetic diffusivity with $\eta_{\rm T}\simeq \eta$, 
a rather small value which is  also known from experiments (Noskov et al. 2012).

\acknowledgements

M. Gellert would like to acknowledge support by the LIMTECH alliance of the Helmholtz association.

\newpage

\end{document}